\documentclass[12pt]{article}
\usepackage{latexsym}
\usepackage{epsfig}
\usepackage{graphicx}
\usepackage[english]{babel}
{\LARGE 
{\large }}

\input colordvi

\newcommand{\be}{\begin{equation}}
\newcommand{\ee}{\end{equation}}
\newcommand{\bea}{\begin{eqnarray}}
\newcommand{\eea}{\end{eqnarray}}\date{\today}
\def\Journal#1#2#3#4{{#1}\, {\bf #2}, #3 (#4)}
  
\def\PLB{Phys.\ Lett.\ B}
\def\PRL{Phys.\ Rev.\ Lett.}

\begin{document}

\title{ Energy losses in the black disc regime and 
correlation effects in the STAR forward pion production in dAu collisions} 

 \author{
Leonid Frankfurt\thanks{frankfur@tauphy.tau.ac.il}\\
Department of Physics, Tel Aviv  University\\
  M. Strikman\thanks{strikman@phys.psu.edu}\\
Department of Physics, Pennsylvania State University,\\ University Park, PA 16802}

\maketitle

\begin{abstract}
We  argue that  in the 
small x processes, in the black disc QCD regime (BDR) a very forward parton propagating through the nuclear 
matter should loose a significant and increasing with energy and atomic 
number fraction of its initial energy as a result of dominance of inelastic 
interactions, causality and energy-momentum conservation. We evaluate these energy losses and find them to  lead to the significant suppression of the 
forward jet production in the central NA collisions at collider energies 
with a moderate suppression of recoiling jet at central rapidities. We 
confront our expectations with the recent RHIC data of  the STAR 
collaboration  on the probability, $P$,  for  emission of at least one 
fast hadron at a central rapidity in association with production of  a  
very forward high $p_t$ neutral pion in $pp$ and $dAu$ collisions.  We 
calculate the A-dependence of  $P$,   and  find that the data imply  a  
strong suppression of leading pion production at  central impact 
parameters.   We also conclude that production of recoil jets in the hard 
subprocess is not suppressed providing further evidence for the dominance 
of peripheral collisions.  Both features of the
data are consistent with the onset of BDR.  We suggest  new phenomena and 
new observables to investigate BDR at RHIC and LHC.

\end{abstract}
\maketitle

\section{Introduction}
\label{intro}
It is well understood now that one of distinctive properties of hard 
processes in pQCD is the fast increase with energy of cross 
sections of  hard  inelastic processes and their significant value. 
Thus the interactions of the partons  produced in the 
sufficiently small x hard processes should be highly inelastic.
Dominance of inelastic processes leads to the specific pattern of 
energy losses for a parton   propagating through the nuclear medium
which is the main subject of this paper. Really in  the  elastic 
rescatterings which dominate in the large x processes energetic parton 
looses a finite energy  ~\cite{Dokshitzer} while propagating a distance $L$:
$\Delta E \approx 0.02 GeV L^2/Fm^2 $.  The analysis of Ref.~\cite{Arleo} 
of the $\pi A$ Drell-Yan pair production indicate that the data are 
consistent with the rate of energy loss by a quark of Ref.~\cite{Dokshitzer} 
and correspond to a energy loss  $\le 4GeV$ for quark  of energy 
$\sim 200 \, GeV$ propagating through a center of a heavy nucleus. 
In contrast in the deep inelastic processes for  example DIS off a proton the
fraction of initial photon energy lost by incident parton is   
$\sim 10\%$ within DGLAP approximations, cf. discussion in section 2. 
Numbers are probably similar within the NLO BFKL approximation 
corresponding to the  rapidity interval between the leading particle 
and next rung in the ladder of about two.
(It is equal to zero  within the  LO BFKL approximation which systematically 
neglects  the loss of energy by energetic particles.)  

In  the black disc regime the contrast between the different patterns of energy 
losses becomes dramatic. 
A parton with energy $E$ propagating
 sufficiently large distance $L$ through
 the nuclear media  should loose energy:
 \begin{equation} 
\Delta E= c E (L /3Fm)
\end{equation}
with $c\approx 0.1$ in small x processes.
This energy loss
exceeds by orders of magnitude  the losses in the large x regime.

Another subtle effect characteristic for a quantum field theory has been found
long before the advent of QCD:  eikonal interactions 
of energetic particle are cancelled out as the consequence of causality 
\cite{Mandelstam, Gribov}.  This  cancellation including additional 
suppression of eikonal diagrams  due to energy-momentum conservation 
is valid for the exchanges by pQCD ladders with vacuum quantum numbers 
in the crossed channel  \cite{BF}. The cancellation of the contribution of 
eikonal diagrams has been demonstrated  also for the exchanges by color 
octet ladders as the consequence of bootstrip condition for the  reggeized gluon 
\cite{BLV}.  Thus sufficiently energetic parton may experience only one 
inelastic collision. To produce $n$ inelastic collisions wave function of energetic 
parton should  develop component containing at least $n$ constituents \cite{BF}. 
This effect leads to the additional depletion of the spectrum of leading partons
in the kinematics close to BDR where inelastic interactions of the energetic parton
is important part of unitarization of amplitudes of hard processes.

Since the number of inelastic collisions  is controlled by the number of scattering 
centres at given impact parameter the effect of the suppression of the yield of leading 
partons should be largest at the central impact parameters.  
We evaluate energy losses of leading parton in small x regime of QCD and  show
that blackening of pQCD interaction leads to dominance of  peripheral 
collisions in the production of the leading hadrons/jets in high 
energy hadron - nucleus interactions and to a significant, increasing with energy and 
atomic number loss of finite fraction of leading parton energy in the central collisions. 
Inclusive cross section is $\propto  A^{1/3}$  deep in the BDR  region with  
suppression of the recoil jets depending on x of jet. One of characteristic features of BDR 
regime is that  there is no suppression of recoil jet in the peripheral collisions.
At moderately small x which are reached  at RHIC, suppression of recoil jet should depend
on its rapidity and be  maximal if  both jets carry a significant fraction 
of the projectile energy.  We  will show that this prediction is supported by the 
recent RHIC data on leading hadron  production in dA collisions.

It is instructive to compare the kinematics of partons involved 
in the production of leading hadrons at RHIC with that for small x phenomena at HERA.
Taking for example the STAR  highest rapidity (y=4)  
and $\left<p_T\right>=1.3 \, GeV/c$ bin \cite{star} we find 
that  $x_N\ge 0.7$  for the incoming parton.   Hence, 
minimal  $x_g$ resolved by such a parton are 
$\sim 4 p_T^2/(x_N s_{NN}) \sim (2 \div 3)\cdot 10^{-4}$.  This is 
very close to the kinematics reached at HERA. The analyses of the HERA 
data within the dipole model approximation show that the partial amplitude 
for the quark interaction reaches at HERA  strength up to  1/2 of the maximal 
strength, see   review in Ref.\cite{annual}.  In the case of heavy nuclei one 
gets an enhancement factor $\sim 0.5 A^{1/3}$  so the quark interaction
with heavy nuclei should be close to BDR for $p_t^2 \le 1.5 GeV^2$ and 
$x_{projectile} \sim 0.5$. In the LHC kinematics  BDR will cover  much 
larger $p_t^2$ range, see for example Fig. 17 in Ref.\cite{annual}.

First evidence for suppression of the forward spectra in the deuteron-gold 
collisions in the kinematics rather close to the BDR was reported by the 
BRAHMS \cite{BRAHMS},  and further studied by PHENIX  \cite{Adler:2004eh}, 
and  STAR\cite{star}. High $p_t$ spectra of  $h^-$  at   
$2\le y \le 3.2 $ are suppressed by a factor
 \cite{BRAHMS} 
\begin{equation} 
R^{h^-}={d\sigma ^{d+A \to h^- +X}\over d y d^2p_T}
\left/ 2A{d\sigma ^{p+p \to h^- +X}\over d y d^2p_T}\right. ,
\end{equation}
which is $\approx 0.8$ for $y=3.2, p_t=2\,\mbox{GeV/c}$.
Since in the kinematics 
of the experiment 
$\sigma(pp\to h^-+X)/\sigma(pp\to h^+ +X)\le 2$,
 the $\pi^-$  yield 
produced by the proton projectile relative to that for the deuteron projectile
(per nucleon) is substantially smaller. As a result in the case of the $\pi^0$ production which is produced with equal strength by protons and neutrons one expects  a bigger suppression. For example   
$R^{\pi^0}_{dAu}(y=3.2, p_t=2\,\mbox{GeV/c}) \approx 0.55$ 
\cite{GSV}).  This suppression factor is significantly larger
than expected suppression due to   the leading twist nuclear 
shadowing. Suppression was observed in the kinematics where 
the hadron production in $pp$ collisions is in a reasonable 
agreement with the recent pQCD calculations based on the 
NLO DGLAP approximation \cite{Werner}. Very recently   
STAR \cite{star} has reported new results for the $\pi^0 $ 
ratios for $y\sim 4$ and $p_t \le 2.0 \, GeV$. They observed a 
larger suppression -  $R^{\pi^0}_{dAu}\sim 1/3$, which  is  
consistent with a  linear extrapolation of  $R^{h^-}_{dAu}$ 
to $y=4$ taking into account the 2/3 factor due to the isospin 
effects \cite{GSV}. 

The STAR experiment also reported the first observation of the correlations 
between the forward  $\pi^0$ production with  the production of the hadrons 
at the central rapidities $|\eta_h|\le 0.75$. Such correlations provide a new 
information about the mechanism of the suppression of the inclusive spectrum. 

The discussion is organized as follows.  In section 2  we evaluate energy 
losses of leading partons of the proton propagating through 
the nuclear medium in the kinematics of the onset of the BDR and find them 
to be $> 10\%$ for central proton-heavy ion collisions in the RHIC kinematics. 
In section 3 we discuss expectations of the QCD BDR for the spectra of the 
leading particles.  In section 4 to disentagle interplay of soft and hard QCD 
phenomena we evaluate correlations between forward  and central hadron 
production using information obtained in the BRAHMS experiment 
\cite{BRAHMS} on the dependence of the central multiplicity on the number of 
the wounded nucleons.   We calculate the dependence of  correlation parameters 
studied by STAR,  on the number of wounded nucleons and find that the data 
require this number to be $\sim 3$, which is significantly smaller than the 
number of  wounded  nucleons  for  central impact parameters $\sim 13$ strongly 
suggesting dominance of the peripheral collisions in $\pi^0$ production.  We also 
want to stress that the paper considers the yield of partons with transverse 
momenta $\le$ than that typical for the BDR. At the same time pion production 
with transverse momenta significantly larger than that typical for BDR should be 
dominated by the scattering at central impact parameters.  For example, Color Glass Condensate (CGC) inspired models predict for this case enhancement of 
production at  central impact parameters by the factor of  
$\approx A^{1/6}$ \cite{Kharzeev}.  In section 5 we  perform a detailed 
analysis of the correlation observables within peripheral models of  pion production 
constrained to  reproduce the inclusive data.
We reproduce the observed values of the correlation parameters
and find that the suppression of the correlation parameter 
related to production of recoil jets observed by STAR \cite{star} 
is due to soft interactions and does not indicate suppression of the pQCD 
mechanism of the production of the recoil jets.  
Thus RHIC data are consistent with the pattern expected  energy losses 
in central collisions,  cf. Ref. \cite{GSV}.  In section 6 we suggest several 
new observables which could allow to diminish model dependence of comparison 
between the hard components of the interaction in $pp$ and $dAu$ cases, 
quantitative study of the suppression on the number of wounded 
nucleons, which also will provide a probe of the color transparency 
effects as well as effects of large gluon fields.

\section{Energy losses of forward parton  in the vicinity of black disk regime}

Energy losses for  the parton propagation through the nucleus medium are dominated in moderate x processes by its elastic  
rescatterings off the constituents of the media due to the Coulomb gluon exchange. Therefore they depend weakly on energy and proportional to  the 
square of distance propagated by the parton \cite{Dokshitzer}.  

However,  the amplitude with color octet quantum numbers  decreases with 
energy due to the gluon reggeization   in pQCD as \cite{LF-Sherman, Lipatov} : 
\begin{equation}
A_g \propto \alpha_s^2 s^{\beta(t)} \left({i+\tan(\pi\beta(t)/2)} \right)
\end{equation}
where $\beta(t)$ is the gluon Regge trajectory with $\beta(t=0) <1$.
Infrared divergences of $\beta(t)$ are regulated by hadron wave functions. 
At the same time the amplitude  due to exchange by a ladder with the 
vacuum quantum numbers  in the crossed channel rapidly grows with energy: 
\begin{equation}
A \propto \alpha_s^2 s^{(1+\lambda(t))} \left(i+\tan((\pi/2)\lambda(t))\right)
\label{ladder}
\end{equation}
where $\lambda(t=0) \approx 0.2$. 
(For the simplicity we restrict ourselves here by the phenomenological fit to 
the theoretical formulae and to the  HERA data on structure functions of a proton.)
Hence such amplitudes (modeled at moderately  small x as the two gluon 
exchange ladder) fastly exceed single gluon exchange term and  at larger 
energies achieve maximum values permitted by probability conservation.

Thus dominance of elastic collisions breaks down at high energies leading 
to the regime where incoherent processes and incoherent energy losses 
dominate leading to the loss of finite fraction of initial energy of a parton, 
cf. Ref.\cite{Martin}. This is the major  difference from moderate x 
processes considered in \cite{Dokshitzer} where coherent energy 
losses seems to  dominate.  Consequently, single inelastic collision of the 
parton produced in a hard high energy NN collision off another nucleon is
described by the  imaginary part of the  two gluon  ladder with the vacuum 
quantum numbers. By definition, the inelastic cross section is calculable in 
terms of  the probability of inelastic interaction,  $P_{inel}(b)$ of a parton 
with a  target at a given impact parameter $b$ \cite{LL}:
\begin{equation}
\sigma_{inel} = \int d^2b P_{inel}(b,s,Q^2)
\label{sb1}
\end{equation}
Since $\sigma_{inel}$ is calculable in QCD \cite{sigma} above equation helps
to calculate $P_{inel}(b,s,Q^2)$ . The probability of inelastic interaction of a 
quark  is cf. \cite{fsw,annual}:  
 \begin{equation}
P_{inel}(b,x,Q^2) = {\pi^2\over 3}\alpha_s(k_t^2) {\Lambda\over k_t^2}xG_A(x,Q^2,b),
\label{sb}
 \end{equation}
where $x\approx 4 k_t^2/s_{qN}, Q^2\approx 4k_t^2,\Lambda \sim 2$ 
(for the gluon case $P_{inel}(b) $ is 9/4 times larger). We use 
gluon density of the nucleus in impact parameter space,
$G_A(x,Q^2,b)$ ($\int d^2b G_A(x,Q^2,b)=G_A(x,Q^2)$) . 
Above equation for the probability of inelastic interaction  is valid 
only for the onset of BDR when $P_{inel}(b,s,Q^2) < 1$ (which is 
the unitarity limit for $P_{inel}(b,s,Q^2) $). 
 
If $P_{inel}(b,x,Q^2) $ as given by Eq.\ref{sb} approaches one or 
exceeds one it means that average number of inelastic interactions, 
$N(b)$ becomes larger than one. Denoting as $ G_{cr} (x,Q^2,b)$  for 
which $P_{inel}(b)$ reaches one we can evaluate  $N(b,x,Q^2)$ as 
\begin{equation}
N(b,x,Q^2)=G_A(x,Q^2,b)/G_{cr} (x,Q^2,b).
\end{equation}
  
As soon as $P_{inel}$ becomes close to one,  we can easily evaluate 
lower boundary for the energy  losses arising from  the single inelastic 
interaction of a parton. This boundary follows from the general properties 
of the parton  ladder. Really, the loss  of 
finite fraction of incident parton energy -$\epsilon$ arises from the 
processes of parton fragmentation into mass $M$ which does not increase 
with energy. For binary collision $M^2={k_t^2\over \epsilon (1-\epsilon)}$. 
For the   contribution of small $\epsilon \le 1/4 $  
\begin{equation}  
\epsilon \approx k_t^2/M^2
\label{fraction}
\end{equation}
Here $k_t$ is transverse momentum of incident parton after inelastic 
collision. The spectrum over the masses in the single ladder approximation 
(NLO DGLAP and BFKL approximations) is  as  follows  
\begin{equation}
d \sigma \propto  \int {dM^2/M^2} (s/M^2)^{\lambda} \theta (M^2-4k_t^2),
\end{equation}
where we accounted for the high energy behavior of the two gluon ladder 
amplitude Eq.(\ref{ladder}). We effectively take into account the energy 
momentum conservation i.e. NLO effects. Consequently the {\it average} 
energy loss (for the contribution of relatively small energy losses 
($\epsilon \le \gamma \sim1/4$) where approximation of Eq.(\ref{fraction}) is valid):
\begin{equation}
\epsilon_N\equiv \left<\epsilon\right> ={\int_0^{\gamma}\epsilon{d\epsilon/\epsilon^{1-\lambda}}
\over \int_0^{\gamma}{d\epsilon/\epsilon^{1-\lambda}}}=\gamma{\lambda\over 1-\lambda}.
\label{epsil}
\end{equation}
For 
the realistic case $\gamma=1/4, \lambda=0.2 $ this calculation gives the fractional energy loss of 6\%. 
This is  lower limit since we neglect here a significant contribution of 
larger $\epsilon$ (it will be calculated elsewhere). 

In the kinematics of onset of BDR  effective number of inelastic interactions becomes 
significantly larger than 1 so in the evaluation of fractional energy loss one should 
multiply evaluated above energy loss by the factor: $N(b)$.  In addition one should 
account for the phenomenon specific for a quantum field theory in small x regime.
The sum of Feynman diagrams which leads to eikonal contribution at moderately 
small  x  is cancelled out at large energies as the consequence of the causality 
i.e. analytic properties of amplitudes and their  decrease with the parton virtuality, 
(cf. Refs. \cite{Mandelstam, Gribov}  and for the generalization to QCD \cite{BLV,BF})
and/or energy-momentum conservation, cf. \cite{annual,BF}.  At central 
impact parameters absorption at high energies  is due to  $N(b)> 1$  
inelastic collisions (interaction with several ladders). The energy of 
initial parton is  shared before collisions at least between N 
constituents in the wave function of the  incident parton to satisfy causality 
and energy-momentum conservation. This quantum field theory effect  
which is absent in the framework of eikonal approximation can be interpreted 
as  an additional energy loss \cite{annual}:
\begin{equation}
\epsilon_{A}(b) \approx N(b)\epsilon_{N}. 
\label{energylosses}
\end{equation} 
Here $\epsilon_{N}$ is the  energy lost due to exchange by one 
ladder - Eq. ~(\ref{epsil}).
Above we do not subtract scattering off nucleon since our interest in the paper 
is in energy losses specific for nuclear processes in the regime when 
interaction with a single nucleon is still far from the BDR.
If collision energies are far from BDR, the energy losses estimated 
above should be multiplied by small probability of secondary 
interactions.  Inclusion  of enhanced "pomeron" diagrams will not change 
our conclusions based on the necessity to account for the energy-momentum conservation law. 

Yields of leading hadrons carrying fraction of projectile momentum $\ge x_F$ are 
rapidly decreasing with $x_N$ as $\propto (1-x_F)^n$. For pion production 
$n\sim 5\div 6$. Obviously for large $x_F$ average values of $x$ 
for progenitor parton are even larger, leading to strong amplification 
of the suppression due to the energy losses. The spectrum of leading pions 
is given in pQCD by the convolution of the quark structure function, 
$\propto (1-x)^n, n\sim 3.5$ and the fragmentation 
function $\propto (1-z)^m, m\sim 1.5 \div 2$ leading to a very steep dependence
on $x_F$, $\propto (1-x_F)^{n+m+1}$. As a result for the STAR kinematics 
$x\sim 0.7$ and $z\sim 0.8$ correspondingly energy losses of 10\% lead to 
a suppression roughly by a factor $[(0.9-x_F)/(1-x_F)]^6$. For $x_F=1/2$ this 
corresponds to suppression by a  factor of four. 
In particular, introducing the energy loss of $\sim 6\%$ 
in the NLO calculation of the pion production is sufficient \cite{GSV}  to 
reproduce the suppression observed by BRAHMS\cite{BRAHMS}.
Similar estimate shows that average losses of $\sim 8\div 10\% $ reproduce 
the suppression of the inclusive yield observed by 
STAR\cite{star}. This value is of the same magnitude as  
the above estimate. Also, Eq.(\ref{energylosses}) leads to much 
stronger suppression for production at  central  impact parameters 
than in peripheral collisions.

In the kinematics of LHC the same $k_t(BDR)$ would be reached 
at  $x_N$ which are smaller by a factor $s_{RHIC}/s_{LHC} \sim 10^{-3}$, 
while for the  same $x_N$ one expects much larger values of $k_t(BDR)$ 
(see e.g. Fig.17  in  \cite{annual}). Thus in the kinematics of LHC 
the regime of large energy losses 
should extend  to smaller $x_N$.

There are two effects associated with  the interaction of  partons in 
the BDR - one is an increase of the transverse momenta of the 
partons and another is the loss of the fraction of the longitudinal 
momentum \cite{Martin}. The net result is that distribution of the 
leading hadrons should drop much stronger with $x_F$  than in the CGC 
models \cite{DGS} where only $k_t$ broadening, change of the resolution 
scale and suppression of coalescence of partons in the final state but 
not the absorption and 
related energy losses were taken into account. At the same time, the  
$k_t$ distribution for  fixed $x_F$ should be broader.   
Note here that the leading particle yield due to the scattering with 
$ k_t \gg  k_{BDR} $ is not suppressed and may give a significant 
contribution at smaller $k_t$ via fragmentation processes.

This discussion shows that selection in the final state of the leading 
hadron ($x_F\ge 0.3\div 0.5$ at RHIC) with moderately large $k_t$ should 
strongly enhance the relative contribution of the peripheral collisions 
where BDR effects are much smaller.  We will demonstrate below that 
these expectations are consistent with the STAR data.

At extremely   high energies where kinematics of the BDR will
be achieved for a broad range of the projectile's parton light-cone 
fractions  and virtualities,  QCD  predicts dominance of scattering 
off the nuclear edge leading to:
\begin{equation}
{d\sigma^{p+A\to \pi + X}\over   d x_N dp_t^2}\left/ \right.
{d\sigma^{p+p\to \pi + X}\over   d x_N dp_t^2}  \propto A^{1/3},
\label{BDR}
\end{equation}
for a large enough $x_N$ and and a wide range of $p_t$. With 
increase of incident energy the range of $p_t$ for fixed $x_N$ would 
increase. Also the suppression for a given $p_t$ would be extended to 
smaller $x_N$.

\section{Interaction  of leading partons with opaque nuclear medium}

At high energies leading partons with light cone momentum $x_N,p_t$ are 
formed before nucleus and can be considered as plane wave if 
\begin{equation}
(x_Ns/m_N)(1/M^2)\gg 2R_A.
\label{planewave}
\end{equation}
Here $M$ is the mass of parton pair (and bremstrahlung gluon) produced  
in the hard collision.  If sufficiently small x are resolved,  the BDR 
regime would be reached:  
\begin{equation}
4 p_t^2/x_N s \le x(BDR).
\label{BDR2}
\end{equation}

In the BDR interaction at impact parameters $b\le R_A$ is strongly  
absorptive as the medium is opaque. As a result, interaction of leading parton lead to a hole of radius $R_A$ in the wave function describing 
incident parton. Correspondingly,  propagation of parton at large impact 
parameters leads to elastic scattering - an analogue of the  Fraunhofer  
diffraction of light off the  black screen.  However since the parton 
belongs to a nucleon, the diffraction for impact parameters larger than 
$R_A+r_{str}$ (where $r_{str}$ is the radius of the strong interaction) 
will lead to the proton in the final state - elastic p A scattering. Only 
for impact parameters $R_A+r_{str} > b > R_A$ the parton may survive to 
emerge in the final state and fragment into the leading hadron. Cross 
section of such diffraction is  $2\pi R_A r_{str}$.  Another contribution 
is due to the propagation of the 
parton through the media. This contribution is suppressed due to 
fractional energy losses which increase with the increase of energy, 
leading to 
gradual decrease of the relative contribution of the inelastic mechanism (see 
discussion in section 5).

Thus we predict that in the kinematics when BDR is achieved in 
pA but not in pN scattering, the  hadron inclusive cross section should be
given by the sum of two terms - scattering from the nucleus
edge which has the same momentum dependence 
as the elementary cross  section and scattering off the opaque media which 
occurs with large energy losses: 
\begin{equation}
{d\sigma(d+A\to h+X)/dx_h d^2p_t \over d\sigma(d+p\to h+X)/dx_h d^2p_t} = 
c_1 A^{1/3} + c_2(A) A^{2/3}
\label{periphery}
\end{equation}
The coefficient $c_1$ is essentially given by the geometry of the nucleus 
edge - cross section for a projectile nucleon to be involved in an inelastic 
interaction with a single nucleon of the target. Coefficient $c_2(A)$ includes 
a  factor due to large energy losses and hence it decreases with increase 
of the incident energy for fixed $x_h, p_t$.  Deep in the BDR the factor 
$c_2(A)$ would  be small enough, so that the periphery term would dominate.

It is worth to compare outlined pattern of interaction in the BDR with 
the expectations of the CGC models for small x hard processes 
in the kinematics where transverse momenta of partons significantly larger
than that characteristic for BDR.  These models employ
the LO BFKL approximation with saturation model \cite{GV} used as 
initial condition of evolution in $\ln(x_o/x)$, see \cite{Dumitru2} and 
references therein. In these models the dependence on atomic number  is 
hidden in the "saturation scale" and in the blackness of interaction at 
this scale. 
In this model partons interact with maximal strength at small 
impact parameters without significant loss  of  energy.
Note that  leading parton looses significant fraction of incident energy  
in the NLO BFKL approximation but not in LO BFKL \cite{NLOBFKL}. 
As a result the cross section is dominated by the scattering at small 
impact parameters and depends on A at energies of RHIC
approximately  as  $A^{5/6}$\cite{Kharzeev}.  Also, the process which 
dominates  in this model at central impact parameters is the scattering 
off the mean field leading (in difference from BDR where DGLAP 
approximation dominates in the peripheral processes in the kinematics of 
RHIC) to events without  balancing jets. With increase of jet transverse 
momenta interaction becomes less opaque, leading to a graduate decrease 
of the probability of inelastic collisions and hence to the dominance of 
the volume term.

 A natural way to distinguish between  these possibilities is to study 
correlations between production of forward high $p_t$ hadrons and 
production of hadrons at central rapidities.  First such study 
was undertaken by the STAR experiment \cite{star}.

  \section{Hadron production in soft nucleon-nucleus interactions at 
central rapidities.}
\label{soft}
The STAR experiment reported correlations between the 
leading pion trigger and central leading charged  hadron production. 
The procedure picks a midrapidity track  with $|\eta_h|\le 0.75$ 
with the highest  $p_T\ge \mbox{0.5 GeV/c}$ and computes the 
azimuthal angle difference $\Delta \phi=\phi_{\pi^0}-\phi_{LCP}$ for 
each event. This provides a coincidence probability $f(\Delta \phi)$.
It is fitted as a sum of two terms - a  background term, $B/2\pi$, 
which is independent of $\Delta \phi$ and the correlation term 
$S(\Delta \phi)$ which is peaked at $\Delta \phi=\pi$. By construction, 
\begin{equation}
\int_0^{2\pi}f(\Delta \phi)d \Delta \phi =B +\int_0^{2\pi}S(\Delta \phi)d \Delta \phi\equiv B+S \le 1.
\label{eq0}
\end{equation}
We will argue below that the A-dependence of $B$ and $S$ is 
sensitive to dependence of the leading pion production
on the centrality of the collision.

The low $p_t$ (soft)  particle contribution which is uncorrelated in 
$\phi$ with the trigger originates both from the collisions of the 
second nucleon of the deuteron with the nucleus and from interactions with 
nucleon involved in the hard collision with several nucleons.  This 
contribution should grow with A since the low $p_t$ hadron multiplicity  
for $y\sim 0$ increases with A. To make quantitative estimate of this 
contribution we will make an approximation that the rate of these soft 
processes is weakly correlated with production of the forward pion 
provided {\it we compare the processes at the same impact parameter}.  
This natural assumption is valid in a  wide range of models including 
CGC models. It is consistent also with the information provided by STAR 
on the weak dependence of the central multiplicity  on $x_F$ of the 
trigger pion, and  lack of long range rapidity correlations for low $p_t$ 
processes which was  observed in many studies of hadron-hadron  collisions.  
Note that at the LHC energies one would have to correct this 
approximation for the correlation of soft and hard interactions in the 
elementary interactions due to more localized transverse distribution of 
the valence partons, see discussion in Ref.\cite{annual}.

Based on generic geometric considerations one expects that the multiplicity 
should be a function of the number of nucleons on the projectile nucleon 
impact parameter. Within this approximation to estimate effects of soft 
production on the correlation observables we can use  information on the 
impact parameter dependence of the hadron multiplicity  which is available 
from several dA RHIC experiments.

 Using the BRAHMS data \cite{BRAHMS} we find that $R^h$  for the STAR 
cuts    can be roughly described by a simple parametrization
 \begin{equation}
R^h=\left({N_{coll}\over 2}\right)^{-r},
\label{2}
\end{equation}
with $r\sim 0.2$. Here  the  factor of two  in the denominator takes 
into account that each of the nucleons of the deuteron experiences, on 
average, equal numbers of collisions.  For example,  for an 
average number of collisions $N_{coll}\approx 7.2$,  Eq.(\ref{2}) 
gives $R^h= 0.77   $  while the BRAHMS data reports 
$R^h= 0.7 \div 0.75 $.

Note in passing, that  the  Gribov-Glauber approximation for the  
hadron - nucleus scattering combined with AGK cutting rules\cite{AGK}   
which neglects energy conservation leads to $R^h=1$. If one takes 
into account energy conservation  - the split of the energy between 
$N_{coll}$, and the increase of the central multiplicity with energy 
$\propto s^{0.2}$  one roughly reproduces Eq. (\ref{2}).

First we want to find out what information about centrality of the interactions 
leading to production of the leading pion   is contained in the A-dependence of  
$B$, the probability that a fast hadron within the experimental cuts does not 
belong to the recoil jet. Obviously, with an increase in the number of   nucleons 
in the nucleus involved in the  interactions practically all events would contain 
at least one particle in the cuts of STAR leading to $B$ very close to one even 
if the elementary hard interaction is not affected by the nuclear environment.  
Using Eq.(\ref{2})    we  can express $B$   for collisions with $n$ nucleons,  $B_n$ 
through characteristics measured for  $pp$ collisions. 

Let us denote the  probabilities to produce a hadron within the central   cuts of 
STAR due to soft and hard interactions in $pp$ collisions by $p_B$ and  $p_S$ 
 respectively. Since the $p_T$ cut of STAR is rather high (comparable 
to the momentum of the leading hadron in the recoiling jet for the 
trigger jet with $\left<p_T\right> \sim \mbox{1.3 GeV/c}$),
we will assume that in the $pp$ events   where both soft and hard 
mechanisms resulted in the   production of a hadron (hadrons) within 
the STAR cuts there is an equal probability  for  the fastest hadron to 
belong to either the soft or  hard component (this is essentially an 
assumption of a reasonably quick convergence of the integrals over 
$p_T$ for $p_{T\, min}=\mbox{0.5 \,  GeV/c}$).
 \footnote{Hereafter we are making an implicit assumption that one can 
neglect production of two hadrons from the soft or hard  pp  interactions 
within the experimental cuts. In the case of soft interactions this is 
justified both by small overall multiplicity 
and presence of short-range negative correlations in rapidities. In 
the case of hard process this is justified by a relatively small value 
of the $p_T$ of the trigger. Obviously one can improve this procedure by 
using information from the STAR experiment
which is  not available yet.}     Within this assumption, 
the probability to produce no fast hadrons is  $(1-p_B)(1-p_S)$; the 
probability to produce a fast   hadron  from the background and 
not from hard process is  $p_B(1-p_S)$; probability to produce a 
fast hadron in hard process and not in the background is   
$p_S(1-p_B)$,   and $p_Sp_B$ is the probability to produce two 
fast hadrons - one in the background and one in the hard process.  Since 
the last outcome contributes equally  to $B_{pp}$ and 
$S_{pp}$ we have
 \begin{equation}
 B_{pp}=p_B (1 - p_S/2),  S_{pp}=p_S(1 -  p_B/2).
 \label{pbps}
 \end{equation}
  Since $S_{pp}$ is small, then  to a very good approximation
   the solution of Eq. ~(\ref{pbps}) is  $p_B=B_{pp}\left(1 +S_{pp}/ (2 - B-S)\right),
   p_S=S_{pp}\left(1 + B_{pp}/ (2 - B-S)\right)$.
Hence  $p_B $ is slightly larger than B, 
while for  $p_S$  a relative correction is significantly larger.

We can now calculate the probability that no hadrons will be 
produced in the inelastic collision of a nucleon  with $m$ 
nucleons of the nucleus:
\begin{equation}
 (1-B-S)_{m\, collisions}=  (1-p_B)^m(1-p_S).
\label{1sb}
 \end{equation}
 
Using STAR data for $S+B$ we find $m= 2.8$.  It is easy to check 
that,  due to  $p_S\ll 1$, this  estimate of $m$ is insensitive to 
the presence of two contributions to the multiplicity.
 
The same picture allows one  to estimate the value of $S$ for $dAu$ collisions. 
Qualitatively,  we expect that $S$ should drop as more hadrons are 
produced in soft collisions and the chance for the fastest hadron 
to be attributed to the recoiling jet becomes smaller. In the case of 
an inelastic collision of a nucleon with  $N$ nucleons of the nucleus, 
the probability that in  exactly $m$ soft interactions a fast hadron 
would be produced, and that also a fast nucleon would be produced in  
a hard collisions is  $p_SC^m_Np_B(1-p_B)^{N-m}$. For these events 
there is $\approx 1/(1+m)$ chance that the fastest  hadron would 
belong to the hard subprocess.
 Summing over $m$ we obtain: 
  \begin{equation}
 S_{N\, collisions}=  p_S\cdot \sum_{m=0}^{m=N} {C_N^m(1-p_B)^{N-m}p_B^m\over (m+1)}. 
 \label{cn}
  \end{equation}
Taking $N\sim 3$ we find $S(dAu)\approx 0.1$ which agrees well with 
the data. Thus we conclude that the increase of the associated 
soft multiplicity  explains the reduction of $S$ observed in the data  
without invoking any suppression of the recoil hadron production on 
the level of the hard subprocess.  
  
We have checked that  accounting for   the decrease of the soft 
multiplicity per $N_{coll}$ leads to a small increase in our estimate 
(see also below).
  
One can see from these equations that if the contribution of the 
central impact parameters ($N_{coll} \sim 13$) were dominating in 
the $\pi^0$ production like in   the CGC models 
one would obtain $(1-B-S),S \ll 0.01$ which is in a qualitative 
contradiction with the data. One would  reach this conclusion  even 
in the case of color transparency for the interaction of  the nucleon
involved in the hard interaction as the second nucleon would 
still experience $\sim 6.5 $ interactions. 
To compare predictions of the CGC models with data 
one should trigger for events at central impact parameters and 
look for suppression of recoil jets in such collisions using the method
described in section \ref{future}.  

Note also that a  simple test of the relative importance   of the 
central and peripheral mechanisms of the pion production is the ratio
of the total hadron  multiplicity in the events with the pion trigger and 
in the minimal bias events. We expect this ratio to be 
$$ (N_{trigger}/N_{min.bias})^{.8}\approx (3/7.2)^{0.8}\approx 1/2.$$
At the same time  in the CGC model this ratio should be larger than one, 
since the relative contribution of the central impact parameters is 
enhanced as compared to the minimal bias sample by a factor 
$A^{1/6} \sim 2.4 $ \cite{Kharzeev}. Unfortunately, information about 
this ratio was not released so far by the STAR collaboration.

 \section{The distribution over the number of  collisions}
 
 In the previous section we calculated $B$ and $S$ for a fixed 
number of collisions.  In a more realistic calculation we need 
to take into account distribution over the number of collisions.  
The important constraint here is that the suppression 
factor,  $R_{dAu}$, for  inclusive $\pi^0$ production is 
$R_{dAu}\sim 0.3$. This requires that  at least nucleons with  
the impact parameter $b\ge b_{min}$ satisfying condition
 \begin{equation}
 \int d^2b T_{A}(b)\theta (b-b_{min}) =R_{dAu},
 \label{16}
 \end{equation}
should contribute to the inclusive pion yield. Here $T_{A}(b)$ is 
the conventional  optical density  which  is expressed through the nuclear 
matter density $\rho_A(r), \int d^3r\rho_A(r)=1$ as 
   \begin{equation}
T_{A}(b) =\int_{-\infty}^{+\infty} dz \rho_A(\sqrt{b^2+z^2}) . 
 \end{equation}
Condition of Eq. (\ref{16}) corresponds  to $b\ge 5\, \mbox{fm}$ for 
$R_{dAu} = 0.3$. For collisions with $b\sim 5\, \mbox{fm}$ the 
average number of collisions is already  larger than 3. However, 
the presence of the more peripheral collisions still may lead to 
an  average number of collisions close to 3.

To simplify the discussion we will consider the case of $pA$ scattering 
and later on correct for the presence of the second  nucleon in 
the projectile. In the probabilistic/geometrical picture 
one can rewrite the inelastic cross section
$\sigma_{in}^{pA}$  as a sum of cross sections with exactly $m$ 
inelastic interactions\cite{BT}:
\begin{equation}
\sigma_{in}^{pA}=\sum_{m=1}^{m=A}\sigma_m, \, \, \sigma_{m}=
{A!\over (A-n)!n!}\int d^2b\,  (T_{A}(b)\sigma_{in}^{NN})^m (1-T_{A}(b)
\sigma_{in}^{NN} )^{A-m}.
\label{sigmam}
\end{equation}
These partial cross sections satisfy the sum rule 
$\sum_{m=1}^{m=A} m \sigma_{m}=A\sigma_{in}^{NN}$.
As a result, if the emission of the particles in each inelastic 
interaction is the same as in the NN collisions, all shadowing  
corrections are canceled reflecting the AGK cancelation \cite{AGK}.

To model distribution over the number of soft interactions, we need 
to introduce a suppression factor SF(b) which is a function of the 
nuclear density per unit area at a given optical density which is given 
by $T(b)$.

Using this model we can calculate the A-dependence of the quantities 
measured by STAR taking into account the distribution over the number of 
the collisions:
\begin{equation}
(1-B-S)= { \sum_{m=1}^{m=A} (1-B-S)_{m\, collisions} \int \sigma_m(b)SF(b) d^2b
\over \sigma_{in}^{NN} R^{\pi^0}_{dAu}},
\end{equation}
where $\sigma_m(b)$ is the integrand of Eq.(\ref{sigmam}) and  
$ (1-B-S)_{m\, collisions} $ is given by Eq.(\ref{1sb}).

Similarly, to calculate $S(pAu)$ we can combine 
Eqs. \ref{cn},\ref{sigmam} to find
\begin{equation}
S= { \sum_{m=1}^{m=A} S_{m\, collisions}\int  \sigma_m(b)SF(b) d^2b
\over \sigma_{in}^{NN} R^{\pi^0}_{dAu}}.
\end{equation}

 For a numerical study we choose two models inspired by the energy 
loss estimate of section 2 for interaction near BDR  and the regime 
of absorption deep in the BDR (cf. Eq. ~(\ref{periphery}))
\footnote{Use of two models allows us to test weak sensitivity of our conclusions to a choice of the specific model for dependence of suppression on the nuclear thickness.}:
\begin{equation} 
SF^{\{1\}}(b)= (1+ a_1 T_{A}(b))^{-1}, \, SF^{\{2\}}(b)= 
(1+ a_2 T_{A}(b))^{-2},
\label{model}
\end{equation}
with the parameter $a_1=2.5, a_2=1.63$ fixed by the condition 
$\int T_{A}(b)SF^i(b) d^2b=R^{\pi^0}_{dAu}=0.286$ as 
measured by STAR for the higher $p_T$ correlation bin 
corresponding to  averaging
 over $30<E_{\pi}<55\, GeV$ \cite{star}.
We found that the two models of suppression give very similar results 
for the observables measured experimentally with the second model 
for $SF(b)$ giving slightly larger values of S and (1-B-S) since it 
yields a stronger suppression of  scattering at the central impact parameters  
(the suppression factor is $\sim 5.4$ in the first model 
and $\sim 20$ in the second model). The numerical values are 
S=0.068  \&  .075;   (1-B-S)=0.070  \& 0.086. If we try to model the 
decrease of the multiplicity per wounded nucleon in line with the BRAHMS 
data we naturally find an increase of $S, (1-B-S)$: 
S=0.085  \&  .090;   (1-B-S)=0.11  \& 0.12.

However,  in the actual experiment the dAu interaction  was 
studied. In this case, the average number of  collisions is about 
a factor 1.4-1.5 higher due to the interaction of the second nucleon.
We can make a rough estimate of  this effect by substituting  
collisions in  pA scattering when the  proton experiences  N inelastic 
interactions by a superposition  with equal probabilities of  N  and 2N 
inelastic collisions.  Clearly, a more detailed modeling of dAu interactions 
is necessary  - we will address it elsewhere. We find, when we   account for 
the energy splitting S=0.067  \&  .072;   
(1-B-S)=0.066  \& 0.079. These numbers should be compared with 
$(1-B-S)=0.1, S=0.093\pm 0.04$ measured by STAR for the higher $p_T$ bin which 
we analyze here.  This suggests that, with inclusion of the second nucleon 
interaction in the Gribov-Glauber model,  one gets a somewhat larger suppression of 
the jets than reported experimentally
\footnote{If some of the pions were due to a production mechanism  without a recoil jet,    $S$ would decrease increasing discrepancy with the data.}
and a smaller probability not 
to observe any hadrons than the one observed experimentally
 leaves  room for  effects due possible deviations
from the geometrical picture. One effect of such kind which was suggested 
in \cite{NP85} is the presence of the color fluctuations the projectile nucleon.
Selection of large $x_F$ in the nucleon may select fluctuations with smaller 
interaction cross section and  lower the number of the 
interactions. However,  the observables used in the analysis are 
not very sensitive to the distribution over the number of interactions 
{\it as long as a peripheral model of $\pi^0$ production is used}.

To summarize, the results of our analysis of the data at higher $p_T$, 
the data are consistent with no suppression of the balancing hadron 
production for the trigger with $\left<p_T \right>\sim 1.3 \, GeV/c$ 
\footnote{In the case of the lower   $p_T$ trigger data set,  our  
estimate for  the dAu scattering gives $(1-B-S)= 0.060-0.072$ and 
$S\sim .045$  which is a bit above the reported value of 
$S=0.020\pm .013$. However application of hard scattering picture in 
any case rather problematic for $p_T(trigger)\sim 1 \, GeV/c$.}. 
Lack of the suppression of the  pQCD mechanism for $<x_A> \sim 0.01$ 
which dominates in the correlation measurements of the STAR puts an upper 
limit on the x range where coherent effects may suppress the pQCD contribution.
Since the analysis of \cite{GSV} find that the pQCD contribution is dominated by
$x_A\ge 0.01$, we can conclude that the main contribution both to inclusive and 
the correlated cross section originates from pQCD hard collisions at large 
impact parameters.

The observation of the recoil jets in the $pp$ case with a strength 
compatible with pQCD calculations suggests that the mechanism for 
pion production in the STAR kinematics is predominantly 
perturbative so that it is legitimate to discuss the propagation of a parton 
through the nucleus leading to pion production.
To ensure a suppression of the pion yield at central impact 
parameters for the discussed kinematics one needs a mechanism 
which is related to the propagation of the projectile parton which is 
generating a pion in a hard interaction with the $x\sim 0.01$ parton.  
For example, the rate of  suppression observed by BRAHMS would 
require fractional energy losses $\sim 3\%$ both in the initial 
and final state \cite{GSV}. Similar losses would produce a suppression 
of the pion yield in  STAR kinematics comparable with the inclusive  data. Modeling 
performed above using Eq.(\ref{model})   indicates 
that  for the central impact parameters the fractional energy losses should be 
at least a factor of 1.5 larger.  Note here that such losses are sufficient only 
because the kinematics of the elementary process 
is close to the limit of the phase space. At the same time, this estimate 
assumes that fluctuations in the energy losses should not be large. 
For example, processes with energy losses comparable to 
the initial energy (like in the case of  high energy electron propagation 
through the media) would not generate necessary suppression provided overall 
losses are of the order of few percent.  
Note also that the second jet in the STAR kinematics 
has much smaller longitudinal momentum and hence is far from the BDR. Therefore 
in the STAR kinematics one does not expect the suppression of the correlation with 
production of the second jet. However a strong suppression is expected for production 
of two balancing forward jets since both of them are interacting in the BDR.

Hence the data are qualitatively consistent with the scenario described in the  
introduction  that leading partons  of the projectile  (with $x\approx 0.7$) interact 
at central impact parameters with the small x nuclear gluon fields with the strength   
close to the BDR.

\section{Suggestions for future measurements to reveal onset of BDR}
\label{future}

We have seen that the quantities used in the STAR analysis involve
an interplay of the hard and soft interactions. Here we want to 
suggest  a few other observables which allow either to suppress 
this interplay or to optimize the  sensitivity to the number of the collisions.

First let us discuss another procedure for studies of the modification 
of the  characteristics of the hard collisions which  is significantly 
less sensitive to the  properties of the soft interactions. Let us 
consider the ratio of the double inclusive and single 
inclusive cross sections for production of a particle in forward 
and in central kinematics which are characterized by their rapidities 
and transverse momenta:
\begin{equation}
RR(y_f,\left|p_{t\,f}\right|,y_c,\left|p_{t\,c}\right|,\phi)=
{d\sigma(y_f,p_{t\,f},y_c,p_{t\,c})\over 
dy_f\,dp_{t\,f}\,dy_c\,dp_{t\,c}}\left/ \right. {d\sigma(y_f,p_{t\,f})\over 
dy_f\,dp_{t\,f}},
\end{equation}
where  $\phi$ is  the angle between$-p_{t\,f}$ and 
$p_{t\,c}$. We can now introduce
 \begin{equation}
\Delta RR(y_f,\left|p_{t\,f}\right|,y_c,\left|p_{t\,c}\right|,\phi)=
RR(y_f,\left|p_{t\,f}\right|,y_c,\left|p_{t\,c}\right|,\phi)-
RR(y_f,\left|p_{t\,f}\right|,y_c,\left|p_{t\,c}\right|,-\phi).
\end{equation}
Similar to the logic of the STAR analysis, we expect that only 
hard contributions to the central production depend on $\phi$.
Hence, in the case of inclusive quantities like $\Delta RR$  the soft 
interactions are  cancelled, while (as we have seen above)
this is not  the case for  the quantities considered in \cite{star}.

Consequently,  the ratio of $\Delta RR^{dAu}$ and $\Delta RR^{pp}$ 
can be used to study how the $p_T$, $\phi$, $\eta$  dependences of the 
balancing jets depend on A (obviously one can consider the ratio of 
$\Delta RR$ integrated over all but one variable).
The STAR analysis used a $p_T\ge 0.5 \, \mbox{GeV/c}$ cutoff to 
enhance the hard contribution. In our procedure one is likely to be 
able to use a smaller cutoff, or no cutoff at all.
It appears that already current statistics of STAR would 
allow at least some of these measurements.  Note here that 
the nuclear shadowing effects are more important for the positive 
rapidities of the recoil jets. Hence,  a study of $\eta$ dependence 
in the kinematics studied by STAR could constrain the leading twist 
shadowing effects between $0.005 < x < 0.02$, albeit for rather 
large impact parameters where shadowing is smaller than on 
average. Note in passing that the current estimates of the 
suppression of the inclusive pion yield due to nuclear shadowing 
overestimate effect as they do not take into account that the process 
is dominated by the scattering at large impact parameters.

It is worth emphasizing here, that for a large range of impact parameters,
one is likely to be in the regime too close to BDR to apply the leading 
twist approximation for  nuclear shadowing. At the same time the 
important feature of the leading twist nuclear shadowing is likely to 
hold, namely that  $R_g=G_A(x,Q^2)/AG_N(x,Q^2)< 1$ is achieved due to 
simultaneous interactions with  $1/R_g$ nucleons, leading to an increase 
in hadron multiplicity at central rapidities and in the nuclear 
fragmentation region \cite{Frankfurt:1998ym}.

To study the dependence of  pion production on the number of 
collisions, one needs to study the multiplicity distribution of the 
soft particle production at central rapidities.
As a first step one would have to deconvolute the hard contribution
which would be well constrained  by a study of $\Delta RR$
(though this is actually a rather small correction, especially 
for large multiplicities). The tail of the distribution at large 
multiplicities  will determine the relative contribution of the 
collisions with several nucleons - for the cuts of STAR the average 
multiplicity should grow $\approx N_{coll}/2$. 

This program would allow for a  study of how the regime of large energy 
losses sets in as a function of the gluon / nucleon transverse density. 
Such a study would have important implications for LHC since  
in the large energy losses scenario, enhancement of the losses 
as compared to LT pQCD calculations is due to the proximity to the unitarity 
limit. Consequently,  one would expect large energy losses for a much 
larger range of rapidities at LHC for the same parton virtualities.

Two  complementary  methods to obtain information about  the 
centrality of dependence of the very forward  pion production would 
be to use information from the zero degree calorimeters (ZDC) along  
the deuteron and gold beams. Measuring the number of neutron spectators  
produced in the fragmentation of the deuteron would be drastically 
different for the peripheral and central impact parameters scenarios - 
in the peripheral case one expects an increase of the spectator 
neutron multiplicity as compared to the minimal bias events as the 
neutron in peripheral interactions has a significant chance to survive 
(provided the pion was emitted in the interaction of the proton 
which occurs with 50\% probability).  We postpone a  quantitative 
description of this significant effect for future publications.
At the same time,  a chance for a 
neutron to escape in the central collisions would be very small: 
$\exp(-\sigma_{in}T(b\sim 0)) \ll 10^{-2}$ \,  \footnote{Since 
the deuteron is weakly bound system, there is a significant tail in 
the wave function at distances $\ge 4 $fm making selection of pure 
central collisions sample very difficult. This problem would be 
greatly alleviated if one would study A-dependence of production of 
spectators with transverse momenta $\ge 200\,  MeV/c$ which 
selects configurations with transverse separations $\le 2 \, fm$.}. The 
ZDC measuring production of neutrons from the nucleus fragmentation 
would observe a smaller than minimal bias multiplicity for the 
peripheral scenario and a larger multiplicity in the central collisions 
scenario. An important advantage of these observables is  that they 
are practically insensitive 
to the issue of the split of the energy between soft interactions. Hence, 
one can reduce uncertainties in the extraction procedure by 
employing information about the production of neutrons in generic 
$dAu$ collisions, in particular in collisions with production of soft 
hadrons at central rapidities. It appears likely that such studies 
would substantially improve the determination of the $T(b)$  
dependence of the suppression factor.

Future experiments at RHIC would allow one  to separate large energy 
losses and  leading twist nuclear shadowing. 
One would have to measure $\Delta RR$ as function of the rapidity. 
The shadowing effects  would lead to drop of  $\Delta RR^{dAu}(\eta)/
\Delta RR^{pp}(\eta)$ for forward rapidities where 
$x_A\sim 10^{-3}$ dominates. Also, if one would be able to decrease  
$x_N$ for fixed small $x_A$, one would enhance the shadowing effects 
as compared to BDR effects. In this limit one would observe an 
increase in  the associated multiplicity at  the central rapidities 
since for the nuclear shadowing mechanism, central impact parameters 
give a large relative contribution.

A color transparency effect would be manifested in a number of 
collisions with small $N_{coll} $ significantly larger than 
that given by the Glauber model. 
Obviously use of the proton beams  would  nicely complement studies 
with the   deuteron beams as one would be able to 
compare triggers for centrality solely based on the interactions of 
one nucleon and on the interactions of two nucleons.

After our first version of our analysis was completed 
PHENIX released the results of their analysis of the correlations 
\cite{Adler:2006hi} which are rather similar to the procedure we 
advocate. They study hadron correlations for smaller rapidities 
and higher $p_t$ than those studied by STAR which is far from the 
BDR and find no suppression of the correlations. 

\section{Conclusions and open questions}

We have demonstrated that partons with large transverse momenta 
corresponding to rather large virtualities for which the BDR is 
reached should lose a substantial fraction of 
their energy.  In the case of inclusive production of very forward pions 
this leads to the dominance of the scattering at peripheral impact 
parameters. For partons with $x_N\ge 0.5$  the  energy losses for 
central impact parameters should lead to suppression of the inclusive 
yield at least by a factor of 
five which corresponds to energy losses $\ge 10\%$.
As a result BDR leads to an extension to a wider $p_t$ range of 
the pattern of the strong suppression of the leading hadron 
production at small $p_t$ observed in the central $pA$ collisions.

With increase of energy from RHIC to LHC energy losses at large $x_N$ 
should strongly increase, while substantial losses $\ge 10\%$  
should persist for rapidities $|y|\ge 2 $. It appears that this 
should lead to increase of the densities in the central collisions 
as compared to the current estimates.  It will also lead to 
suppression of the production of the recoil jets at the rapidity 
intervals where no suppression is present at RHIC.  In the forward 
direction we expect a significantly larger suppression than 
already large  suppression found in \cite{DGS} where fractional 
energy losses were neglected.  Fractional energy losses result in modification of the form of the QCD factorization theorem at LHC energies. In particular they lead to suppression of Higgs / SUSY particle production in pp scattering by at least $\left[(1-x/(1+\epsilon))/(1-x)\right]^{10}$. Here $x\ge m_H/\sqrt{s}$ are the light-cone fractions carried by initial gluons which initiate production of the Higgs particle with accompanying bremstrahlung,  and $\epsilon \ge .1$ is the fractional energy loss. This corresponds to a suppression $ \ge \, 10 \%$ for $m_H=140 $ GeV.

 Further studies of the   proton/deuteron - nucleus interactions 
at the central impact parameters at RHIC and future experiments at 
LHC  would provide important constrains on this important ingredient  
of high energy dynamics. Similar effects will be present in the central 
$pp$ collisions at LHC. They would amplify the correlations between the 
hadron production in the fragmentation and central regions 
discussed  in Ref.\cite{annual}.

We would like to thank Yu.Dokshitzer, S.~Heppelmann, D.~Kharzeev, L.~McLerran, G.~Rakness, 
R.~Venugopalan, I.~Vitev and W.~Vogelsang for useful discussions. The 
work was supported by the DOE and BSF grants.


\begin{thebibliography}{10}

\bibitem{Dokshitzer}
   R.~Baier, Y.~L.~Dokshitzer, A.~H.~Mueller, S.~Peigne and D.~Schiff,
  %``Radiative energy loss of high energy quarks and gluons in a  finite-volume
  %quark-gluon plasma,''
  Nucl.\ Phys.\ B {\bf 483}, 291 (1997)

%%%%%%%%%%
\bibitem{Arleo} F.Arleo 
%Constraints on quark energy loss from Drell-Yan data.
Phys.Lett.B532,231-239,2002
%%%%%%%%%%
Lett.\  {\bf 85}, 5535 (2000)


\bibitem{Mandelstam}
  S.~Mandelstam,
  %``Cuts In The Angular Momentum Plane. 2,''
  Nuovo Cim.\  {\bf 30}, 1148 (1963). 

\bibitem{Gribov} V.N. Gribov,  
The Theory of complex angular momenta: Gribov lectures on theoretical physics,
Cambridge Univ. Press, 2003.

\bibitem{BF} B.~Blok and L.~Frankfurt,
   ``The casuality and/or energy-momentum conservation constraints on QCD
  amplitudes in small x regime,'' arXiv:hep-ph/0611062. 

\bibitem{BLV} J.Bartels, L.Lipatov and G.Vacca 
Nucl.Phys.B706(2005)391 


\bibitem{star}
 J.~Adams {\it et al.}  [STAR Collaboration],
  %``Forward neutral pion production in p+p and d+Au collisions at s(NN)**(1/2)
  %= 200-GeV),''
  arXiv:nucl-ex/0602011.

\bibitem{annual} L.~Frankfurt, M.~Strikman and C.~Weiss,
Annu. Rev. Nucl. Part. Sci.  {\bf 55}, 403 (2005). 
hep-ph/0507286.

\bibitem{BRAHMS}   I.~Arsene, {\it et al.},
             \Journal{\PRL}{93}{242303}{2004};
           {\bf 91}, 072305 (2003).

           \bibitem{Adler:2004eh}
  S.~S.~Adler {\it et al.}  [PHENIX Collaboration],
  %``Nuclear modification factors for hadrons at forward and backward
  %rapidities in deuteron gold collisions at s(NN)**(1/2) = 200-GeV,''
  Phys.\ Rev.\ Lett.\  {\bf 94}, 082302 (2005)
 % [arXiv:nucl-ex/0411054].
  
\bibitem{GSV}
           V.~Guzey, M.~Strikman, and W.~Vogelsang,
             \Journal{\PLB}{603}{173}{2004}.  arXiv:nucl-ex/0602011.

\bibitem{Werner} F.~Aversa, P.~Chiappetta, M.~Greco and J.~P.~Guillet,
  %``QCD Corrections To Parton-Parton Scattering Processes,''
  Nucl.\ Phys.\ B {\bf 327}, 105 (1989);  B.~Jager, A.~Schafer, M.~Stratmann and W.~Vogelsang,
  %``Next-to-leading order QCD corrections to high-p(T) pion production in
  %longitudinally polarized p p collisions,''
  Phys.\ Rev.\ D {\bf 67}, 054005 (2003)
%  [arXiv:hep-ph/0211007];  D.~de Florian,
  %``Next-to-leading order QCD corrections to one hadron production inpolarized
  %p p collisions at RHIC,''
  Phys.\ Rev.\ D {\bf 67}, 054004 (2003)
 % [arXiv:hep-ph/0210442].



\bibitem{Kharzeev}  D.~Kharzeev, E.~Levin and L.~McLerran,
  %``Parton saturation and N(part) scaling of semi-hard processes in QCD,''
  Phys.\ Lett.\ B {\bf 561}, 93 (2003)
%  [arXiv:hep-ph/0210332];
  
  
D.~Kharzeev, Y.~V.~Kovchegov and K.~Tuchin,
  %``Cronin effect and high-p(T) suppression in p A collisions,''
  Phys.\ Rev.\ D {\bf 68}, 094013 (2003)
%  [arXiv:hep-ph/0307037].
        
         \bibitem{LF-Sherman}L.~L.~Frankfurt and V.~E.~Sherman,
   %``Reggeization Of Vector Meson And Vacuum Singularity In Renormalizable
  %Yang-Mills Models,''
  Sov.\ J.\ Nucl.\ Phys.\  {\bf 23}, 581 (1976).
 
 \bibitem{Lipatov} L.N. Lipatov, Sov. J. Nucl. Phys. 23 (1976) 338


    
\bibitem{Martin} 
  L.~Frankfurt, V.~Guzey, M.~McDermott and M.~Strikman,
  %``Revealing the black body regime of small x DIS through final state
  %signals,''
 Phys.\ Rev.\ Lett.\  {\bf 87}, 192301 (2001)
  [arXiv:hep-ph/0104154].
  
 L.~Frankfurt and M.~Strikman,
  %``Ion induced quark-gluon implosion,''
Phys.\ Rev.\ Lett.\  {\bf 91}, 022301 (2003)
 [arXiv:nucl-th/0212094].
  
    
\bibitem{LL} L.Landau and E.Lifshitz , "Nonrelativistic Quantum Mechanics"
Pergamon Press.

\bibitem{sigma} 
B.~Blaettel, G.~Baym, L.~L.~Frankfurt, H.~Heiselberg and M.~Strikman,
  %``Hadronic cross-section fluctuations,''
  Phys.\ Rev.\ D {\bf 47}, 2761 (1993).

L.~Frankfurt, G.~A.~Miller and M.~Strikman,
  %``Coherent nuclear diffractive production of mini - jets: Illuminating color
  %transparency,''
  Phys.\ Lett.\ B {\bf 304}, 1 (1993)
  [arXiv:hep-ph/9305228].


L.~Frankfurt, A.~Radyushkin and M.~Strikman,
  %``Interaction of small size wave packet with hadron target,''
  Phys.\ Rev.\ D {\bf 55}, 98 (1997)
  [arXiv:hep-ph/9610274].


\bibitem{fsw}
  L.~Frankfurt, M.~Strikman and C.~Weiss,
  %``Dijet production as a centrality trigger for p p collisions at LHC,''
  Phys.\ Rev.\ D {\bf 69}, 114010 (2004)
  [arXiv:hep-ph/0311231].

\bibitem{DGS}
A.~Dumitru, L.~Gerland and M.~Strikman,
  %``Proton breakup in high-energy p A collisions from perturbative QCD,''
  Phys.\ Rev.\ Lett.\  {\bf 90}, 092301 (2003)
  [Erratum-ibid.\  {\bf 91}, 259901 (2003)]
%  [arXiv:hep-ph/0211324].
\bibitem{GV}K.~Golec-Biernat and M.~Wusthoff,
%   ``Saturation effects in deep inelastic scattering at low Q**2 and its
  %implications on diffraction,''
Phys.\ Rev.\ D {\bf 59}, 014017 (1999)
  [arXiv:hep-ph/9807513].

\bibitem{Dumitru2} A.~Dumitru, A.~Hayashigaki and J.~Jalilian-Marian,
%   ``Geometric scaling violations in the central rapidity region of d + Au
  %collisions at RHIC,''
  Nucl.\ Phys.\ A {\bf 770}, 57 (2006)
  [arXiv:hep-ph/0512129].

\bibitem{NLOBFKL} M.~Ciafaloni and G.~Camici,
  %``Energy scale(s) and next-to-leading BFKL equation,''
  Phys.\ Lett.\ B {\bf 430}, 349 (1998)
%  [arXiv:hep-ph/9803389].
        
M.~Ciafaloni and D.~Colferai,
  %``The BFKL equation at next-to-leading level and beyond,''
Phys.\ Lett.\ B {\bf 452}, 372 (1999)
[arXiv:hep-ph/9812366].
  
V.~S.~Fadin and L.~N.~Lipatov,
  %``BFKL pomeron in the next-to-leading approximation,''
Phys.\ Lett.\ B {\bf 429}, 127 (1998)
[arXiv:hep-ph/9802290].

\bibitem{AGK} V. A. Abramovsky, V. N. Gribov, and O. V. Kancheli, 
  Yad. Fiz. 18, 595 (1973) [Sov. J. Nucl. Phys. 18, 308 (1974)]. 
  

\bibitem{BT}  L.~Bertocchi and D.~Treleani,
  %``Glauber Theory, Unitarity, And The Agk Cancellation,''
  J.\ Phys.\ G {\bf 3}, 147 (1977).

\bibitem{NP85}L.~L.~Frankfurt and M.~I.~Strikman,
  %``Point - Like Configurations In Hadrons And Nuclei And Deep Inelastic
  %Reactions With Leptons: EMC And EMC Like Effects,''
  Nucl.\ Phys.\ B {\bf 250}, 143 (1985).
  
 \bibitem{Frankfurt:1998ym}
  L.~Frankfurt and M.~Strikman,
  %``Diffraction at HERA, color opacity and nuclear shadowing,''
  Eur.\ Phys.\ J.\ A {\bf 5}, 293 (1999)
  [arXiv:hep-ph/9812322].
  
       \bibitem{Adler:2006hi}
  S.~S.~Adler {\it et al.}  [PHENIX Collaboration],
  %``Azimuthal angle correlations for rapidity separated hadron pairs in d + Au
  %collisions at s(NN)**(1/2) = 200-GeV,''
[arXiv:nucl-ex/0603017].


 \end{thebibliography}
\end{document}